\documentclass[sigconf,10pt,nonacm]{acmart}

\AtBeginDocument{%
  }

\usepackage{amsmath}
\usepackage{soul}

\begin{document}
\title{GPUs, CPUs, and… NICs: Rethinking the Network's Role in Serving Complex AI Pipelines
}
\author{Mike Wong}
\email{mikedwong@cs.princeton.edu}
\affiliation{%
  \institution{Princeton University}
  \author{}
  \country{}
}
\author{Ulysses Butler}
\email{ulysses.butler@nyu.edu}
\affiliation{%
  \institution{New York University}
  \author{}
  \country{}
}
\author{Emma Farkash}
\email{ef2830@princeton.edu}
\affiliation{%
  \institution{Princeton University}
  \author{}
  \country{}
}
\author{Praveen Tammana}
\email{praveent@cse.iith.ac.in}
\affiliation{%
  \institution{Indian Institute of Technology Hyderabad}
  \author{}
  \country{}
}

\author{Anirudh Sivaraman}
\email{anirudh@cs.nyu.edu}
\affiliation{%
  \institution{New York University}
  \author{}
  \country{}
}
\author{Ravi Netravali}
\email{rnetravali@cs.princeton.edu}
\affiliation{%
  \institution{Princeton University}
  \author{}
  \country{}
}

\pagestyle{plain}
\newcommand{\mw}[1] {{\textcolor{cyan}{MW: {#1}}}}
\newcommand{\rn}[1] {{\textcolor{red}{RN: {#1}}}}
\newcommand{\as}[1] {{\textcolor{blue}{AS: {#1}}}}
\newcommand{\uab}[1] {{\textcolor{purple}{UAB: {#1}}}}

\newcommand{\ie}{{i.e.,}~}
\newcommand{\eg}{{e.g.,}~}
\newcommand{\aka}{{a.k.a.,}~}
\newcommand{\para}[1]{\medskip\noindent\textbf{#1}}
\newcommand{\paraf}[1]{\noindent\textbf{#1}}

\newcommand{\cmark}{\ding{51}}
\newcommand{\xmark}{\ding{55}}
\newcommand\comb[2][^n]{\prescript{#1\mkern-0.5mu}{}C_{#2}}
\newenvironment{parafont}{\fontfamily{ptm}\selectfont}{}
\newcommand{\Para}[1]{\vspace{2pt}\noindent\begin{parafont}\textbf{\textit{#1}}\end{parafont}}
\begin{abstract}


The increasing prominence of AI necessitates the deployment of inference platforms for efficient and effective management of AI pipelines and compute resources. As these pipelines grow in complexity, the demand for distributed serving rises and introduces much-dreaded network delays. In this paper, we investigate how the network can instead be a boon to the excessively high resource overheads of AI pipelines. To alleviate these overheads, we discuss how resource-intensive data processing tasks -- a key facet of growing AI pipeline complexity -- are well-matched for the computational characteristics of packet processing pipelines and how they can be offloaded onto SmartNICs. We explore the challenges and opportunities of offloading, and propose a research agenda for integrating network hardware into AI pipelines, unlocking new opportunities for optimization.

\end{abstract}

\maketitle

\section{Introduction}




 AI is now used to guide or automate decisions in applications across societal sectors, with use cases in healthcare, finance, Internet services, and more \cite{the_evolution_of_drones,cartel, campus-shuttle,are-we-ready-for-ai-powered-security-cameras, medicalexample, economicsllm}. 
The systems community has played a large role in realizing this revolution. Indeed, the backbone of this success is the inference platforms that support running ML models, coordinating user requests, and managing compute resources~\cite{clipper, clockwork, nexus, shepherd, pagedattention}.


The most recent wave of AI innovation has brought pipelines whose increased complexity heightens the need for \emph{distributed serving}.
First, models and request rates continue to grow and often cannot be supported by single nodes/GPUs, motivating efforts to parallelize~\cite{distrifusion, megatron, alpa} (e.g., tensor parallelism) or disaggregate tasks (e.g., LLM prefill and decode) across nodes~\cite{distserve, splitwise, mooncake}.
Second, the number of inference iterations required to respond to a single request is increasing -- e.g., denoising for image generation~\cite{distrifusion, nirvana} or inference-time scaling~\cite{inference-time} -- necessitating distribution to parallelize inference steps for acceptable latency. 
Lastly, compound AI~\cite{cacheblend, ragcache, ralm, ragserve}, and agentic workflows~\cite{parrot, swarm, autogpt, claude} that fuse together multiple models and non-parametric components (e.g., vector databases, verifiers) are routinely scheduled across multiple nodes given the varying resource needs of constituent components.

In this paper, we argue that the push towards distributed serving actually presents new opportunities for networks to aid resource-constrained serving platforms, rather than serving as solely a headache with transmission delays. In particular, while programmable network hardware (e.g., SmartNICs, which is the focus of this work) is ill-suited for costly model inference, the growing number of surrounding data processing tasks in emerging pipelines presents a more favorable picture (\S\ref{sec:motivation}). These tasks, which can push CPU/GPU utilization to nearly 100\% (\S\ref{sec:motivation}), are responsible for transforming, formatting, and filtering data between and within model inferences, e.g.,  image normalization, tokenization, and KV cache compression. Our main observation is that such tasks often consist of a bounded number of deterministic steps and rely on large, but simple, data structures -- attributes that make them well-suited for the highly parallel nature of SmartNIC packet processing pipelines. Moreover, the close proximity of SmartNICs to traditional CPU/GPU servers ensures on-path processing while data is in transit, without introducing new hops.



Capitalizing on this promise, however, raises several questions: Which functions are amenable to network hardware offload? How can such tasks be packaged to fit within the tight compute and resource constraints of SmartNICs, without harming fidelity or slowing down network traffic? We investigate these key questions and motivate a research agenda around solving them in the following sections: \S\ref{sec:motivation} taxonomizes AI pipeline tasks and identifies those suitable for offload; \S\ref{sec:challenges}-\ref{s:offloading} build on this, presenting fundamental challenges with implementing offloads, as well as pathways forward (with implementation sketches for three concrete examples); \S\ref{sec:related-work} outlines related works.
Finally, \S\ref{sec:roadmap} outlines open questions and a future agenda. Our hope is to motivate the systems community to integrate broader, heterogeneous resources into the purview of AI serving runtimes, motivating new lines of work in distributed scheduling, cross-platform task compilation, and more.

\section{What to Offload and Why?}
\label{sec:motivation}

Model serving involves more than solely invoking neural networks; serving tasks also include detailed logging, management of hardware resources, pre/post-processing of data and results, maintaining (or transferring) intermediate state, loading models, and more. We explore the resource needs of such tasks, and identify those that are amenable to (and worthwile for) network hardware offload below.

\para{Memory management.} This class of tasks focus on managing where and how models and internal state are stored. These techniques often either coordinate and change how memory is used for neural networks on GPUs or use host memory to extend the limited memory available on GPUs \cite{pagedattention, ragserve}. Due to the distance from the GPUs and the low memory available on network accelerators, memory management is not deemed a viable candidate for offload.

\para{Scheduling and batching.} Serving systems embed algorithms to determine when and how the inference engine is invoked on inputs \cite{orca,sarathi-serve,preeble}. Scheduling techniques are often difficult to parallelize because these functions involve conditional logic (e.g., sorting requests and predicting memory/compute overheads), which creates dependencies that force sequential execution, leading to resource underutilization on network accelerators.
%

\para{Model execution.} The execution of the neural network inference on incoming data inputs. Neural networks incur high memory overheads that far exceed the memory capacities of network hardware. Moreover, serving platforms are typically already equipped with GPUs and other accelerators that have already been specially designed and heavily optimized for the compute tasks in model execution.

    

    

In contrast to these jobs, we find that the \textbf{data processing} operations which transform, filter, and format data are most suitable for offload. 
These tasks can run both around model inferences (e.g., preparing data as input to a model such as tokenization or image normalization) or within a model inference (e.g., compressing KV cache state prior to transmission in distributed LLM serving~\cite{distserve}).
Importantly, such functions can be executed \emph{inline} while intermediate data is in transit between two nodes and do not require additional transfers to the SmartNIC.
Moreover, many of these functions employ a bounded number of deterministic steps and rely on large, but simple, data structures, making them amenable to the pipeline and data parallelism that is fundamental to network hardware.
For instance, image normalization ingests a large matrix containing and independently applies subtraction and division operations to each pixel value.
Similarly, KV cache compression algorithms often process large tensors by multiplying their values by a scaling factor for quantization.


\para{Potential benefits.} 
To understand the value of offloading data processing tasks, we profiled resource usage for several representative AI pipelines that encompass key emerging paradigms including multi-model agents (both language and vision) and various distributed ML serving paradigms \cite{llava, flamingo, imagebind, clip}. All experiments were run with the TorchServe serving platform~\cite{torch-serve} across two 32-core 64-hyperthread 256-GB DRAM machine (AMD EPYC 7543P) with 4 NVIDIA A6000 GPUs. 
Across these pipelines, we observe median and maximum CPU utilization values of 86\% and 98\%, respectively, even when GPUs are fully utilized for model inference. Crucially, were data processing tasks to be offloaded to SmartNICs, these values could be drastically reduced by 32\% and 70\%, drastically improving end-to-end response latencies that prior work has show to suffer from such resource overheads and contention~\cite{goldminer,scheduling_overhead,tf-data}.


\section{Challenges of SmartNIC Offloading}
\label{sec:challenges}
To offload a program onto a network accelerator, we first use a specification of the data processing function.
This function ingests a data sample and applies transformations to modify it.
We seek to generate a semantically equivalent implementation of this function that runs on the SmartNIC. As an example, one popular data processing function is \texttt{Resize}, which performs weighted averages of an image's input pixels to resize it for downstream CNN processing.
However, generating such an implementation is not straightforward. 
It must realize the behavior of the specification while processing one packet of data at a time rather than the entire data sample all at once.
Moreover, the implementation is subject to a number of constraints such as packet size, resource limits (e.g., BRAM), and SLOs.
In this section, we dive deeper into these challenges and discuss avenues to overcome them.

\subsection{Finite Packet Size}
\para{Challenge.} 
Packet processing pipelines don’t have access to the full data context all at once.
Since data is divided into packets, only one chunk of data can be processed at a time. 
E.g., Ethernet frames can be up to 1500 bytes in size, whereas AXI transactions, commonly used in switches and FPGAs, are typically 32 bytes.
This is especially problematic as text prompts and images get larger and larger.
One option is to buffer all incoming packets together within the device to reconstruct the entire data sample.
However, this would significantly reduce throughput because the buffering would stall the pipeline, preventing output packets from being generated until the full data sample is assembled and processed. 

\para{Opportunity.}
We observe that although data processing functions are implemented with the assumption that the full data context is available, the sub-operations that are performed only require operating on localized "windows" of the input (e.g., neighboring pixels in an image or nearby words/characters in a sentence) at a time to maintain semantic equivalence.
That is, operations don't typically require data from distant segments of the input.
Intuitively, localized windows of data have high levels of correlation and are often processed and aggregated together.
We should thus only buffer enough packets for each localized data window before generating output packets rather than waiting for the entire data context to be reconstructed.
To minimize packet buffering, we look to modify the serialization on the data source to change the data arrangement.


\subsection{Compute Overheads}
\para{Challenge.} 
Data processing operations that perform series of arithmetic transformations often require compute-intensive operations that are difficult to run on network accelerators.
E.g., it is common for models to require input normalization which involves division and floating point operations.
Some network devices do not support these computations while others can but at the cost of high resource overheads and significantly reduced packet processing performance.

\para{Opportunity.}
A big contributor to the abstraction gap between the data processing specification and the capabilities of network hardware is that many specified operations are difficult to run directly on the device.
Network hardware is much better suited for parallel rule-based pattern matching and is why they have excelled in traditional networking tasks such as packet switching, intrusion detection, and header parsing \cite{intrusiondetect, packetswitch}.
Expressing data processing operations in this matching paradigm is key to bridging this gap.

We observe that specific operations and constant operands for compute-intensive operations are often known a priori and do not change dynamically.
This property is present in image processing functions that perform deterministic arithmetic computations on pixel values and text replacements for LLM prompts.
Because of this observation, we do not need to directly implement arithmetic operations on the device; instead, we can use memoization to store lookup tables that map each input value in the data sample to the pre-computed result.




\subsection{Limited Memory}
\para{Challenge.} 
Network accelerators often have limited on-chip memory; this has a big impact when a pipeline requires employing many data processing functions in tandem or when a function requires excessively large lookup tables.
FPGA-based SmartNICs in particular often only have several tens of MB of high-speed BRAM available which makes it difficult to store look-up tables, packet data, tokenizer vocabularies, and other requisite application data.
Although some FPGAs feature on-board DRAM and HBM, read and write operations with these memory types incur significantly higher latencies. 
Additionally, they offer limited support for parallel lookups compared to BRAM.
Given BRAM's characteristics, its usage is strongly encouraged for optimal performance.

\para{Opportunity.}
As prior works have observed, data across requests and packets exhibit a significant degree of similarity.
In video data, neighboring frames within the same temporal window will have similar content \cite{twostream}.
Within a frame, the pixel distribution is often skewed -- our analysis of the Kinetics video dataset \cite{kinetics} reveals that the top 10 pixels consume appear almost 20\% of the time.
In LLMs, users edit and refine prompts causing them to contain a lot of redundant text \cite{paceprompts}.
To perform more complex tasks, users or applications often include highly similar contexts with input prompts that include domain knowledge \cite{cachegen,preeble}.

Due to this redundancy in data content, we can use off-chip memories (e.g., DRAM and HBM for FPGAs) for the less frequently accessed table entries while using faster on-chip memory (e.g., BRAM) for the entries most relevant to current requests. 
For instance, a lookup table that cannot fit entirely in BRAM can move less frequently accessed table entries to other, more distant memory types.
In doing so, large lookupt tables can still be employed while maximizing the number of BRAM accesses, thereby significantly reducing the number of cycles needed to perform reads and writes.

\subsection{Parallel Processing}
\para{Challenge.}
While network accelerators are much more energy efficient compared to traditional processors, they also have much lower clock speeds.
To maximize performance wins and best leverage the hardware capabilities, it is imperative to employ data and pipeline parallelism.
However, naively parallelizing can introduce additional resource overheads or harm correctness if done haphazardly.

\para{Opportunity.}
Data processing tasks often consist of a fixed number of compute steps with little to no dependencies and operate on large data structures. 
Images are represented using large matrices and require a sequence of transformations to modify its size, content, and structure.
Textual prompts are composed of large vectors of characters that require transformations for normalization and tokenization.
As a result, they are amenable to the pipeline and data parallelism that is fundamental to network hardware.

\section{Example Offloads}
\label{s:offloading}

In this section, we exemplify the challenges and solution directions from \S\ref{sec:challenges} using three widely-used example data processing tasks. 

\subsection{Image Normalization}
The \texttt{Normalize} function subtracts each pixel value by the mean of its channel and then divides it by the standard deviation of its channel.
We denote $\textbf{I}$ and $\textbf{O}$ as the input and output images respectively and $c$ as the color channel.

\[ \textbf{O}[c,i,j] = \frac{\textbf{I}[c,i,j] - mean[c]} {std[c]} \]

Note that mean and standard deviation for each channel are normally provided a priori and do not need to be computed.

\para{Compute overheads.}
To reduce compute overheads, our design uses one lookup table for each color channel.
Each table will contain 255 entries, with each entry mapping a color intensity value to the pre-computed normalized value.
This results in significant compute savings as a lookup and a write typically incur 1 clock cycle for each operation whereas division can consume many 10s of clock cycles for just a single operation.


\subsection{Bilinear Interpolation} \label{sec:5.2}
Bilinear interpolation is a popular algorithm used for many image processing functions such as resizing, rotating, and texture mapping. 
For each output pixel, it finds four corresponding pixels in the input image and computes the following result.

\[ 
\begin{split}
\textbf{O}[x,y] = (1 - \Delta_x)(1-\Delta_y)\textbf{I}[x_1, y_1] + \\
\Delta_x(1-\Delta_y)\textbf{I}[x_2, y_1] + \\
(1 - \Delta_x)\Delta_y\textbf{I}[x_1, y_2] + \\
\Delta_x\Delta_y\textbf{I}[x_2, y_2]   
\end{split}
\]

\para{Processing with packets.}
Instead of buffering the entire image, we only need to buffer one additional row of pixel data within the image to store enough neighboring pixels for the localized window.
A single row for a 1280 x 720 image requires 3840 bytes, or 120 AXI transmissions (each consumes 32 bytes), and the first output pixel can be generated once those packets have been stored.
Once the first row of output pixels has been generated, the process is repeated for the remainder of the image.
As a result, the pipeline is only stalled for the duration of one row before the current row of output packets can start being generated.

\para{Serialization.}
Serialization can reduce buffering significantly. 
The image can be first divided into `tiles` where each tile contains the input pixels for a single output pixel.
The bigger the discrepancy between the input and output image sizes, the more distant the pixels in each tile.
Instead of storing the image on a row-by-row basis, all pixels within a tile can be stored adjacent to each other. 
When the tile-major image is processed by the network accelerator, it can immediately begin computing output pixels without waiting for subsequent packets.

\noindent
\textbf{Row-major order}: \[ I[0,0], I[0,1], I[0,2], I[0,3], I[0,4], ... \] 

\noindent
\textbf{Tile-major order}: \[  I[0,0], I[0,1], I[1,0], I[1,1], I[0,2], ... \]

For larger discrepancies, pixels in the input image may be completely skipped over and discarded. 
Since the image sizes are known a priori, customized serialization can exclude those pixels from being stored and transmitted in the first place, reducing transmission overheads.


\para{Compute overheads.}
We pre-compute lookup tables instead of performing all arithmetic operations directly on the device.
We assume a target image size of 224 x 224, which is commonly used by many vision models \cite{resnet, vgg, ImageNet}.
However, this is not straightforward at first glance.
Each output pixel requires first finding the four corresponding neighboring pixels in the input image and then using the interpolation formula above.

We first observe that the four input pixel positions as well as $\Delta_x$ and $\Delta_y$ can be pre-computed offline since they are only dependent on the size of the output image, which remains constant.
Moreover, $\Delta_x$ and $\Delta_y$ have only 224 different possible values each.
But pre-computing all results of the entire formula would require an excessively large amount of memory because it would necessitate one entry for every $\Delta_x$, $\Delta_y$, and 4-pixel `tile` ($224^2 \times 255^4$ total combinations).
Instead, we pre-compute each term in the interpolation formula rather than the full result.
There are $224 \times 224$ combinations of $\Delta_x$ and $\Delta_y$, 4 different terms (each using a different combination of $\Delta_x$ and $\Delta_y$), and 255 pixels resulting in approximately 51M lookup entries.
This saves significant compute overheads by avoiding the expensive arithmetic operations involved in finding the input pixels and performing interpolation by only requiring the addition of the four pre-computed terms.

\para{Memory.}
Each table entry consists of 34 bits: 2 bits for the tile index (to codify which of the four interpolation terms to retrieve), 16 bits for the $(x,y)$ coordinates of the output image, 8 bits for the input pixel, and 8 bits for the pre-computed value.
A table with 51M 34-bit entries requires about 216MB of memory.
We observe that because the order of the output pixel generation is predictable, we can simply keep the relevant table entries for the part of the image that is currently being processed in BRAM and gradually swap them out for entries in later parts of the image as the function progresses.
Further savings can be reaped by leveraging the high degree of similarity across frames and within individual images to retain the most relevant pixel color entries in BRAM.

\subsection{Text Processing}

Tokenization is an essential function for LLMs that maps an input text prompt into a series of tokens. 
Below is an example input prompt and its tokenized sequence.

\vspace{1em}
\noindent
\textbf{Prompt}: This is an example of an input prompt

\noindent
\textbf{Tokens}: ["This", " is", " an", " exam", "ple", " of", " an", " input", " prom", "pt"]

\para{Packet processing.}
Tokenization is challenging when processing one chunk of data at a time because the input prompt will be split at every 32 bytes, which is problematic because tokens will be interrupted.
Our solution is to prepend the last several characters of each chunk to the chunk in the following packet so that there is textual overlap between neighboring chunks.
The amount of characters stored is equal to the maximum token length.
When two neighboring chunks are tokenized, the same token will appear twice surrounding any tokens that were incorrectly generated. 
Once these are removed, we get the correct tokenized sequence.
Below, brackets denote the additional text that was added from the previous chunk.

\noindent
\textbf{Prompt in chunks}: ["This is an e", "[ an e]xample of an", "[ of an]input prompt"] 

\noindent
\textbf{Tokens}: ["This", " is", " an", " e", " an", "exam", "ple", " of", " an", " input", " prom", "pt"] \\
\textbf{Corrected tokens}: ["This", " is", \st{" an"}, \st{" e"}, " an", "exam", "ple", \st{" of"}, \st{" an"}, " of", " an", " input", " prom", "pt"]

We observe that parts of the packet metadata remain largely unchanged or are unneeded for data processing functions.
This space can be repurposed to hold the additional tokens until they are de-duplicated at the end of the pipeline.

\para{Parallelism.}
Although each chunk requires serial processing because  token lengths are unknown a priori, our design lets us reap high degrees of pipeline parallelism.
We first allocate different pipeline stages for text normalization and tokenization.
Within tokenization, we further parallelize by having each  stage perform several matches on its respective chunk before it is sent to the next stage.
This is a huge benefit over existing implementations that run serially.
There are other attempts at parallelizing tokenization but they require the entire data context first to properly split the text without disrupting token boundaries (e.g., splitting by periods) \cite{youtokentome}.

\subsection{Additional Examples}
Our approach is broadly applicable to other tasks and domains. 
Images and tabular datasets necessitate estimating missing or corrupted values through imputation, which often involves using statistical measures such as the mean, median, or mode of neighboring values \cite{impute}.
Moreover, real-time monitoring systems frequently perform data cleaning operations to filter out anomalous sensor readings \cite{iot}. 
Further, feature encoding converts raw categorical data into numerical representations \cite{tabular}.
We believe many of these operations can be realized into a semantically equivalent SmartNIC implementation by employing serialization to alter how the data is stored and by pre-computing results in lookup tables.

\section{Roadmap: Automatic Compilation of SmartNIC Offloads} 
\label{sec:roadmap}

As AI pipelines become more and more complex, seamless and efficient offload tools will free AI developers from the burden of worrying about the underlying details of these smart network hardware devices. Moving forward, our goal is to automate the generation of data processing tasks for SmartNICs, capitalizing on the potential to alleviate resource tensions illustrated above. This requires addressing several key questions such as: Can we build a compiler that successfully maps a data processing specification to a semantically equivalent implementation that runs within the within resource constraints? How do we support a diverse set of different data processing operations and hardware backends? 

At a higher level, we envision such a compiler to run alongside updated distributed scheduling engines in AI serving runtimes. As done today, such platforms can ingest graph-like specifications of the AI pipeline to run, outlining both the requisite components and the data/control-level coordination between them~\cite{parrot}. From there, our proposed compiler would be responsible for analyzing the pipeline along two dimensions to produce a task layout across available hardware: (1) determining what to offload based on end-to-end bottleneck analysis of individual requests, i.e., tasks that are truly on-path (and would not introduce unnecessary communication overheads if offloaded) and would alleviate (rather than introduce) slowdowns, and (2) how to offload, i.e., addressing the compilation questions above and integrating insights akin to those in \S\ref{sec:challenges}-\ref{s:offloading} but in an automated fashion.


\section{Related Work}
\label{sec:related-work}


\para{Reducing CPU overheads.} Several works acknowledge and alleviate CPU bottlenecks in ML workloads and focus on resource scaling, profiling, and pre-processing acceleration of \emph{training pipelines} \cite{goldminer,where-is-my-training-bottleneck,data-stalls-ml-pipeline,tf-data,cachew}. In contrast, our focus is on \emph{inference pipelines} that differ in 2 ways: (1) they have strict SLO requirements, and (2) they feature multiple ML models with their associated tasks, rather than a single model. DALI \cite{dali} is a library for offloading pre-processing operations onto GPUs; instead, owing to GPU oversubscriptions common at the edge~\cite{ekya,mu}, we target offloading pipeline components to non-traditional (network) accelerators for AI.

\para{Network hardware offload.} Several works have leveraged network hardware for executing machine learning models~\cite{switches-dream-of-ml,taurus,homunculus,n3ic,network-is-ai-accelerator} or performing gradient aggregation~\cite{netreduce,atp} in networks. Instead of offloading model inference itself as these prior works do, we focus on offloading a growing list of CPU-based tasks that sit in the interstices of model inference calls in vision pipelines. 

\bibliographystyle{ACM-Reference-Format}
\bibliography{references}


\begin{thebibliography}{64}


\ifx \showCODEN    \undefined \def \showCODEN     #1{\unskip}     \fi
\ifx \showDOI      \undefined \def \showDOI       #1{#1}\fi
\ifx \showISBNx    \undefined \def \showISBNx     #1{\unskip}     \fi
\ifx \showISBNxiii \undefined \def \showISBNxiii  #1{\unskip}     \fi
\ifx \showISSN     \undefined \def \showISSN      #1{\unskip}     \fi
\ifx \showLCCN     \undefined \def \showLCCN      #1{\unskip}     \fi
\ifx \shownote     \undefined \def \shownote      #1{#1}          \fi
\ifx \showarticletitle \undefined \def \showarticletitle #1{#1}   \fi
\ifx \showURL      \undefined \def \showURL       {\relax}        \fi
\providecommand\bibfield[2]{#2}
\providecommand\bibinfo[2]{#2}
\providecommand\natexlab[1]{#1}
\providecommand\showeprint[2][]{arXiv:#2}

\bibitem[sch({[n.\,d.]})]%
        {scheduling_overhead}
 \bibinfo{year}{[n.\,d.]}\natexlab{}.
\newblock \bibinfo{title}{{Can Scheduling Overhead Dominate LLM Inference Performance? A Study of CPU Scheduling Overhead on Two Popular LLM Inference Systems}}.
\newblock
\newblock
\shownote{\url{https://mlsys.wuklab.io/posts/scheduling_overhead/}, Retrieved on 2025-01}.


\bibitem[swa({[n.\,d.]})]%
        {swarm}
 \bibinfo{year}{[n.\,d.]}\natexlab{}.
\newblock \bibinfo{title}{{Swarm}}.
\newblock
\newblock
\shownote{\url{https://github.com/openai/swarm/tree/main /}, Retrieved on 2025-01}.


\bibitem[the({[n.\,d.]})]%
        {the_evolution_of_drones}
 \bibinfo{year}{[n.\,d.]}\natexlab{}.
\newblock \bibinfo{title}{{The Evolution of Drone Technology: Embracing AI and Computer Vision in 2024}}.
\newblock
\newblock
\shownote{\url{https://visionplatform.ai/computer-vision-for-drones-and-uav-in-2024}, Retrieved on 2024-06}.


\bibitem[you({[n.\,d.]})]%
        {youtokentome}
 \bibinfo{year}{[n.\,d.]}\natexlab{}.
\newblock \bibinfo{title}{{YouTokenToMe Text Tokenizer}}.
\newblock
\newblock
\shownote{\url{https://github.com/vkcom/youtokentome /}, Retrieved on 2025-01}.


\bibitem[dal(2024)]%
        {dali}
 \bibinfo{year}{2024}\natexlab{}.
\newblock \bibinfo{title}{{NVIDIA Data Loading Library (DALI)}}.
\newblock \bibinfo{howpublished}{\url{https://github.com/NVIDIA/DALI}}.
\newblock


\bibitem[{ Alexander Isenko, Ruben Mayer, Jeffrey Jedele, Hans-Arno Jacobsen }(2022)]%
        {where-is-my-training-bottleneck}
\bibfield{author}{\bibinfo{person}{{ Alexander Isenko, Ruben Mayer, Jeffrey Jedele, Hans-Arno Jacobsen }}.} \bibinfo{year}{2022}\natexlab{}.
\newblock \showarticletitle{{Where is my training bottleneck? hidden trade-offs in deep learning preprocessing pipelines}}. In \bibinfo{booktitle}{\emph{SIGMOD}}.
\newblock


\bibitem[{ Jayashree Mohan, Amar Phanishayee, Ashish Raniwala, Vijay Chidambaram }(2021)]%
        {data-stalls-ml-pipeline}
\bibfield{author}{\bibinfo{person}{{ Jayashree Mohan, Amar Phanishayee, Ashish Raniwala, Vijay Chidambaram }}.} \bibinfo{year}{2021}\natexlab{}.
\newblock \showarticletitle{{Analyzing and mitigating data stalls in dnn training}}. In \bibinfo{booktitle}{\emph{VLDB}}.
\newblock


\bibitem[Agarwal et~al\mbox{.}(2023)]%
        {nirvana}
\bibfield{author}{\bibinfo{person}{Shubham Agarwal}, \bibinfo{person}{Subrata Mitra}, \bibinfo{person}{Sarthak Chakraborty}, \bibinfo{person}{Srikrishna Karanam}, \bibinfo{person}{Koyel Mukherjee}, {and} \bibinfo{person}{Shiv Saini}.} \bibinfo{year}{2023}\natexlab{}.
\newblock \bibinfo{title}{Approximate Caching for Efficiently Serving Diffusion Models}.
\newblock
\showeprint[arxiv]{2312.04429}~[cs.CV]
\urldef\tempurl%
\url{https://arxiv.org/abs/2312.04429}
\showURL{%
\tempurl}


\bibitem[Alayrac et~al\mbox{.}(2022)]%
        {flamingo}
\bibfield{author}{\bibinfo{person}{Jean-Baptiste Alayrac}, \bibinfo{person}{Jeff Donahue}, \bibinfo{person}{Pauline Luc}, \bibinfo{person}{Antoine Miech}, \bibinfo{person}{Iain Barr}, \bibinfo{person}{Yana Hasson}, \bibinfo{person}{Karel Lenc}, \bibinfo{person}{Arthur Mensch}, \bibinfo{person}{Katie Millican}, \bibinfo{person}{Malcolm Reynolds}, \bibinfo{person}{Roman Ring}, \bibinfo{person}{Eliza Rutherford}, \bibinfo{person}{Serkan Cabi}, \bibinfo{person}{Tengda Han}, \bibinfo{person}{Zhitao Gong}, \bibinfo{person}{Sina Samangooei}, \bibinfo{person}{Marianne Monteiro}, \bibinfo{person}{Jacob Menick}, \bibinfo{person}{Sebastian Borgeaud}, \bibinfo{person}{Andrew Brock}, \bibinfo{person}{Aida Nematzadeh}, \bibinfo{person}{Sahand Sharifzadeh}, \bibinfo{person}{Mikolaj Binkowski}, \bibinfo{person}{Ricardo Barreira}, \bibinfo{person}{Oriol Vinyals}, \bibinfo{person}{Andrew Zisserman}, {and} \bibinfo{person}{Karen Simonyan}.} \bibinfo{year}{2022}\natexlab{}.
\newblock \bibinfo{title}{Flamingo: a Visual Language Model for Few-Shot Learning}.
\newblock
\showeprint[arxiv]{2204.14198}~[cs.CV]
\urldef\tempurl%
\url{https://arxiv.org/abs/2204.14198}
\showURL{%
\tempurl}


\bibitem[{Amey Agrawal, Nitin Kedia, Ashish Panwar, Jayashree Mohan, Nipun Kwatra, Bhargav Gulavani, Alexey Tumanov, Ramachandran Ramjee}(2024)]%
        {sarathi-serve}
\bibfield{author}{\bibinfo{person}{{Amey Agrawal, Nitin Kedia, Ashish Panwar, Jayashree Mohan, Nipun Kwatra, Bhargav Gulavani, Alexey Tumanov, Ramachandran Ramjee}}.} \bibinfo{year}{2024}\natexlab{}.
\newblock \bibinfo{title}{{Taming Throughput-Latency Tradeoff in LLM Inference with Sarathi-Serve}}.
\newblock


\bibitem[Bai et~al\mbox{.}(2022)]%
        {claude}
\bibfield{author}{\bibinfo{person}{Yuntao Bai}, \bibinfo{person}{Saurav Kadavath}, \bibinfo{person}{Sandipan Kundu}, \bibinfo{person}{Amanda Askell}, \bibinfo{person}{Jackson Kernion}, \bibinfo{person}{Andy Jones}, \bibinfo{person}{Anna Chen}, \bibinfo{person}{Anna Goldie}, \bibinfo{person}{Azalia Mirhoseini}, \bibinfo{person}{Cameron McKinnon}, \bibinfo{person}{Carol Chen}, \bibinfo{person}{Catherine Olsson}, \bibinfo{person}{Christopher Olah}, \bibinfo{person}{Danny Hernandez}, \bibinfo{person}{Dawn Drain}, \bibinfo{person}{Deep Ganguli}, \bibinfo{person}{Dustin Li}, \bibinfo{person}{Eli Tran-Johnson}, \bibinfo{person}{Ethan Perez}, \bibinfo{person}{Jamie Kerr}, \bibinfo{person}{Jared Mueller}, \bibinfo{person}{Jeffrey Ladish}, \bibinfo{person}{Joshua Landau}, \bibinfo{person}{Kamal Ndousse}, \bibinfo{person}{Kamile Lukosuite}, \bibinfo{person}{Liane Lovitt}, \bibinfo{person}{Michael Sellitto}, \bibinfo{person}{Nelson Elhage}, \bibinfo{person}{Nicholas Schiefer}, \bibinfo{person}{Noemi Mercado},
  \bibinfo{person}{Nova DasSarma}, \bibinfo{person}{Robert Lasenby}, \bibinfo{person}{Robin Larson}, \bibinfo{person}{Sam Ringer}, \bibinfo{person}{Scott Johnston}, \bibinfo{person}{Shauna Kravec}, \bibinfo{person}{Sheer~El Showk}, \bibinfo{person}{Stanislav Fort}, \bibinfo{person}{Tamera Lanham}, \bibinfo{person}{Timothy Telleen-Lawton}, \bibinfo{person}{Tom Conerly}, \bibinfo{person}{Tom Henighan}, \bibinfo{person}{Tristan Hume}, \bibinfo{person}{Samuel~R. Bowman}, \bibinfo{person}{Zac Hatfield-Dodds}, \bibinfo{person}{Ben Mann}, \bibinfo{person}{Dario Amodei}, \bibinfo{person}{Nicholas Joseph}, \bibinfo{person}{Sam McCandlish}, \bibinfo{person}{Tom Brown}, {and} \bibinfo{person}{Jared Kaplan}.} \bibinfo{year}{2022}\natexlab{}.
\newblock \bibinfo{title}{Constitutional AI: Harmlessness from AI Feedback}.
\newblock
\showeprint[arxiv]{2212.08073}~[cs.CL]
\urldef\tempurl%
\url{https://arxiv.org/abs/2212.08073}
\showURL{%
\tempurl}


\bibitem[{Bret Hull, Vladimir Bychkovsky, Yang Zhang, Kevin Chen, Michel Goraczko, Allen Miu, Eugene Shih, Hari Balakrishnan, Samuel Madden}(2006)]%
        {cartel}
\bibfield{author}{\bibinfo{person}{{Bret Hull, Vladimir Bychkovsky, Yang Zhang, Kevin Chen, Michel Goraczko, Allen Miu, Eugene Shih, Hari Balakrishnan, Samuel Madden}}.} \bibinfo{year}{2006}\natexlab{}.
\newblock \showarticletitle{{CarTel: A Distributed Mobile Sensor Computing System}}. In \bibinfo{booktitle}{\emph{SenSys}}.
\newblock


\bibitem[Cassel({[n.\,d.]})]%
        {are-we-ready-for-ai-powered-security-cameras}
\bibfield{author}{\bibinfo{person}{David Cassel}.} \bibinfo{year}{[n.\,d.]}\natexlab{}.
\newblock \bibinfo{title}{Are We Ready for AI-Powered Security Cameras?}
\newblock \bibinfo{howpublished}{\url{https://thenewstack.io/are-we-ready-for-ai-powered-security-cameras/}}.
\newblock


\bibitem[Chatterjee and Ahmed(2022)]%
        {iot}
\bibfield{author}{\bibinfo{person}{Ayan Chatterjee} {and} \bibinfo{person}{Bestoun~S. Ahmed}.} \bibinfo{year}{2022}\natexlab{}.
\newblock \showarticletitle{IoT anomaly detection methods and applications: A survey}.
\newblock \bibinfo{journal}{\emph{Journal of King Saud University - Computer and Information Sciences}} \bibinfo{volume}{34}, \bibinfo{number}{8} (\bibinfo{year}{2022}), \bibinfo{pages}{4547--4563}.
\newblock
\urldef\tempurl%
\url{https://doi.org/10.1016/j.jksuci.2022.01.001}
\showDOI{\tempurl}


\bibitem[Crankshaw et~al\mbox{.}(2017)]%
        {clipper}
\bibfield{author}{\bibinfo{person}{Daniel Crankshaw}, \bibinfo{person}{Xin Wang}, \bibinfo{person}{Giulio Zhou}, \bibinfo{person}{Michael~J. Franklin}, \bibinfo{person}{Joseph~E. Gonzalez}, {and} \bibinfo{person}{Ion Stoica}.} \bibinfo{year}{2017}\natexlab{}.
\newblock \bibinfo{title}{Clipper: A Low-Latency Online Prediction Serving System}.
\newblock
\showeprint[arxiv]{1612.03079}~[cs.DC]
\urldef\tempurl%
\url{https://arxiv.org/abs/1612.03079}
\showURL{%
\tempurl}


\bibitem[{Dan Graur, Damien Aymon, Dan Kluser, Tanguy Albrici, Chandramohan A Thekkath, and Ana Klimovic }(2022)]%
        {cachew}
\bibfield{author}{\bibinfo{person}{{Dan Graur, Damien Aymon, Dan Kluser, Tanguy Albrici, Chandramohan A Thekkath, and Ana Klimovic }}.} \bibinfo{year}{2022}\natexlab{}.
\newblock \showarticletitle{{Cachew: Machine learning input data processing as a service. }}. In \bibinfo{booktitle}{\emph{USENIX ATC}}.
\newblock


\bibitem[{Derek G. Murray, Jiří Šimša, Ana Klimovic, and Ihor Indyk}(2021)]%
        {tf-data}
\bibfield{author}{\bibinfo{person}{{Derek G. Murray, Jiří Šimša, Ana Klimovic, and Ihor Indyk}}.} \bibinfo{year}{2021}\natexlab{}.
\newblock \showarticletitle{{tf.data: A Machine Learning Data Processing Framework}}. In \bibinfo{booktitle}{\emph{Proc. VLDB Endow. 14, 12}}.
\newblock


\bibitem[Dong et~al\mbox{.}(2024)]%
        {paceprompts}
\bibfield{author}{\bibinfo{person}{Yihong Dong}, \bibinfo{person}{Kangcheng Luo}, \bibinfo{person}{Xue Jiang}, \bibinfo{person}{Zhi Jin}, {and} \bibinfo{person}{Ge Li}.} \bibinfo{year}{2024}\natexlab{}.
\newblock \bibinfo{title}{PACE: Improving Prompt with Actor-Critic Editing for Large Language Model}.
\newblock
\showeprint[arxiv]{2308.10088}~[cs.CL]
\urldef\tempurl%
\url{https://arxiv.org/abs/2308.10088}
\showURL{%
\tempurl}


\bibitem[Fu et~al\mbox{.}(2024)]%
        {inference-time}
\bibfield{author}{\bibinfo{person}{Yichao Fu}, \bibinfo{person}{Junda Chen}, \bibinfo{person}{Siqi Zhu}, \bibinfo{person}{Zheyu Fu}, \bibinfo{person}{Zhongdongming Dai}, \bibinfo{person}{Aurick Qiao}, {and} \bibinfo{person}{Hao Zhang}.} \bibinfo{year}{2024}\natexlab{}.
\newblock \bibinfo{title}{Efficiently Serving LLM Reasoning Programs with Certaindex}.
\newblock
\showeprint[arxiv]{2412.20993}~[cs.LG]
\urldef\tempurl%
\url{https://arxiv.org/abs/2412.20993}
\showURL{%
\tempurl}


\bibitem[Gao et~al\mbox{.}(2024)]%
        {pagedattention}
\bibfield{author}{\bibinfo{person}{Yi Gao}, \bibinfo{person}{Xue Li}, \bibinfo{person}{Xinyu Chen}, \bibinfo{person}{Xian Liu}, \bibinfo{person}{Jun Song}, \bibinfo{person}{Liang Zeng}, \bibinfo{person}{Shaojie Zhang}, \bibinfo{person}{Haoyu Chen}, \bibinfo{person}{Wei Huang}, {and} \bibinfo{person}{Zhiqiang Wu}.} \bibinfo{year}{2024}\natexlab{}.
\newblock \showarticletitle{Efficient Memory Management for Large Language Model Serving with PagedAttention}.
\newblock \bibinfo{journal}{\emph{ACM Transactions on Computing Systems (TOCS)}} \bibinfo{volume}{42}, \bibinfo{number}{3} (\bibinfo{year}{2024}), \bibinfo{pages}{1--25}.
\newblock
\urldef\tempurl%
\url{https://doi.org/10.1145/3600006.3613165}
\showDOI{\tempurl}


\bibitem[Gerbasi et~al\mbox{.}(2023)]%
        {medicalexample}
\bibfield{author}{\bibinfo{person}{A. Gerbasi}, \bibinfo{person}{G. Clementi}, \bibinfo{person}{F. Corsi}, {and} \bibinfo{person}{et al.}} \bibinfo{year}{2023}\natexlab{}.
\newblock \showarticletitle{DeepMiCa: Automatic segmentation and classification of breast MIcroCAlcifications from mammograms}.
\newblock \bibinfo{journal}{\emph{Computational Methods and Programs in Biomedicine}}  \bibinfo{volume}{235} (\bibinfo{year}{2023}), \bibinfo{pages}{107483}.
\newblock
\urldef\tempurl%
\url{https://doi.org/10.1016/j.cmpb.2023.107483}
\showDOI{\tempurl}


\bibitem[Girdhar et~al\mbox{.}(2023)]%
        {imagebind}
\bibfield{author}{\bibinfo{person}{Rohit Girdhar}, \bibinfo{person}{Alaaeldin El-Nouby}, \bibinfo{person}{Zhuang Liu}, \bibinfo{person}{Mannat Singh}, \bibinfo{person}{Kalyan~Vasudev Alwala}, \bibinfo{person}{Armand Joulin}, {and} \bibinfo{person}{Ishan Misra}.} \bibinfo{year}{2023}\natexlab{}.
\newblock \bibinfo{title}{ImageBind: One Embedding Space To Bind Them All}.
\newblock
\showeprint[arxiv]{2305.05665}~[cs.CV]
\urldef\tempurl%
\url{https://arxiv.org/abs/2305.05665}
\showURL{%
\tempurl}


\bibitem[Gujarati et~al\mbox{.}(2020)]%
        {clockwork}
\bibfield{author}{\bibinfo{person}{Arpan Gujarati}, \bibinfo{person}{Reza Karimi}, \bibinfo{person}{Safya Alzayat}, \bibinfo{person}{Wei Hao}, \bibinfo{person}{Antoine Kaufmann}, \bibinfo{person}{Ymir Vigfusson}, {and} \bibinfo{person}{Jonathan Mace}.} \bibinfo{year}{2020}\natexlab{}.
\newblock \showarticletitle{Serving DNNs like Clockwork: Performance Predictability from the Bottom Up}. In \bibinfo{booktitle}{\emph{14th {USENIX} Symposium on Operating Systems Design and Implementation ({OSDI} 20)}}. \bibinfo{publisher}{{USENIX} Association}, \bibinfo{pages}{443--462}.
\newblock
\showISBNx{978-1-939133-19-9}
\urldef\tempurl%
\url{https://www.usenix.org/conference/osdi20/presentation/gujarati}
\showURL{%
\tempurl}


\bibitem[{Gyeong-In Yu,Joo Seong Jeong, Geon-Woo Kim, Soojeong Kim, Byung-Gon Chun}(2022)]%
        {orca}
\bibfield{author}{\bibinfo{person}{{Gyeong-In Yu,Joo Seong Jeong, Geon-Woo Kim, Soojeong Kim, Byung-Gon Chun}}.} \bibinfo{year}{2022}\natexlab{}.
\newblock \bibinfo{title}{{Orca: A Distributed Serving System for Transformer-Based Generative Models}}.
\newblock


\bibitem[{Hanyu Zhao, Zhi Yang, Yu Cheng, Chao Tian, Shiru Ren, Wencong Xiao, Man Yuan, Langshi Chen, Kaibo Liu, Yang Zhang, Yong Li, and Wei Lin}(2023)]%
        {goldminer}
\bibfield{author}{\bibinfo{person}{{Hanyu Zhao, Zhi Yang, Yu Cheng, Chao Tian, Shiru Ren, Wencong Xiao, Man Yuan, Langshi Chen, Kaibo Liu, Yang Zhang, Yong Li, and Wei Lin}}.} \bibinfo{year}{2023}\natexlab{}.
\newblock \showarticletitle{{Goldminer: Elastic scaling of training data pre-processing pipelines for deep learning}}. In \bibinfo{booktitle}{\emph{Proc. ACM Manag. Data, 1(2)}}.
\newblock


\bibitem[He et~al\mbox{.}(2015)]%
        {resnet}
\bibfield{author}{\bibinfo{person}{Kaiming He}, \bibinfo{person}{Xiangyu Zhang}, \bibinfo{person}{Shaoqing Ren}, {and} \bibinfo{person}{Jian Sun}.} \bibinfo{year}{2015}\natexlab{}.
\newblock \bibinfo{title}{Deep Residual Learning for Image Recognition}.
\newblock
\urldef\tempurl%
\url{https://doi.org/10.48550/ARXIV.1512.03385}
\showDOI{\tempurl}


\bibitem[Hutchings et~al\mbox{.}(2002)]%
        {intrusiondetect}
\bibfield{author}{\bibinfo{person}{B.L. Hutchings}, \bibinfo{person}{R. Franklin}, {and} \bibinfo{person}{D. Carver}.} \bibinfo{year}{2002}\natexlab{}.
\newblock \showarticletitle{Assisting network intrusion detection with reconfigurable hardware}. In \bibinfo{booktitle}{\emph{Proceedings. 10th Annual IEEE Symposium on Field-Programmable Custom Computing Machines}}. \bibinfo{pages}{111--120}.
\newblock
\urldef\tempurl%
\url{https://doi.org/10.1109/FPGA.2002.1106666}
\showDOI{\tempurl}


\bibitem[Jin et~al\mbox{.}(2024)]%
        {ragcache}
\bibfield{author}{\bibinfo{person}{Chao Jin}, \bibinfo{person}{Zili Zhang}, \bibinfo{person}{Xuanlin Jiang}, \bibinfo{person}{Fangyue Liu}, \bibinfo{person}{Xin Liu}, \bibinfo{person}{Xuanzhe Liu}, {and} \bibinfo{person}{Xin Jin}.} \bibinfo{year}{2024}\natexlab{}.
\newblock \bibinfo{title}{RAGCache: Efficient Knowledge Caching for Retrieval-Augmented Generation}.
\newblock
\showeprint[arxiv]{2404.12457}~[cs.DC]
\urldef\tempurl%
\url{https://arxiv.org/abs/2404.12457}
\showURL{%
\tempurl}


\bibitem[Kay et~al\mbox{.}(2017)]%
        {kinetics}
\bibfield{author}{\bibinfo{person}{Will Kay}, \bibinfo{person}{Joao Carreira}, \bibinfo{person}{Karen Simonyan}, \bibinfo{person}{Brian Zhang}, \bibinfo{person}{Chloe Hillier}, \bibinfo{person}{Sudheendra Vijayanarasimhan}, \bibinfo{person}{Fabio Viola}, \bibinfo{person}{Tim Green}, \bibinfo{person}{Trevor Back}, \bibinfo{person}{Paul Natsev}, \bibinfo{person}{Mustafa Suleyman}, {and} \bibinfo{person}{Andrew Zisserman}.} \bibinfo{year}{2017}\natexlab{}.
\newblock \bibinfo{title}{The Kinetics Human Action Video Dataset}.
\newblock
\showeprint[arxiv]{1705.06950}~[cs.CV]
\urldef\tempurl%
\url{https://arxiv.org/abs/1705.06950}
\showURL{%
\tempurl}


\bibitem[Korinek(2023)]%
        {economicsllm}
\bibfield{author}{\bibinfo{person}{Anton Korinek}.} \bibinfo{year}{2023}\natexlab{}.
\newblock \bibinfo{title}{The Role of Large Language Models in Automating Economic Research}.
\newblock
\urldef\tempurl%
\url{https://www.nber.org/system/files/working_papers/w30957/w30957.pdf}
\showURL{%
\tempurl}
\newblock
\shownote{Working Paper}.


\bibitem[Lao et~al\mbox{.}(2021)]%
        {atp}
\bibfield{author}{\bibinfo{person}{ChonLam Lao}, \bibinfo{person}{Yanfang Le}, \bibinfo{person}{Kshiteej Mahajan}, \bibinfo{person}{Yixi Chen}, \bibinfo{person}{Wenfei Wu}, \bibinfo{person}{Aditya Akella}, {and} \bibinfo{person}{Michael Swift}.} \bibinfo{year}{2021}\natexlab{}.
\newblock \showarticletitle{{ATP}: In-network Aggregation for Multi-tenant Learning}. In \bibinfo{booktitle}{\emph{18th USENIX Symposium on Networked Systems Design and Implementation (NSDI 21)}}. \bibinfo{publisher}{USENIX Association}, \bibinfo{pages}{741--761}.
\newblock
\showISBNx{978-1-939133-21-2}
\urldef\tempurl%
\url{https://www.usenix.org/conference/nsdi21/presentation/lao}
\showURL{%
\tempurl}


\bibitem[Li et~al\mbox{.}(2024)]%
        {distrifusion}
\bibfield{author}{\bibinfo{person}{Muyang Li}, \bibinfo{person}{Tianle Cai}, \bibinfo{person}{Jiaxin Cao}, \bibinfo{person}{Qinsheng Zhang}, \bibinfo{person}{Han Cai}, \bibinfo{person}{Junjie Bai}, \bibinfo{person}{Yangqing Jia}, \bibinfo{person}{Ming-Yu Liu}, \bibinfo{person}{Kai Li}, {and} \bibinfo{person}{Song Han}.} \bibinfo{year}{2024}\natexlab{}.
\newblock \bibinfo{title}{DistriFusion: Distributed Parallel Inference for High-Resolution Diffusion Models}.
\newblock
\showeprint[arxiv]{2402.19481}~[cs.CV]
\urldef\tempurl%
\url{https://arxiv.org/abs/2402.19481}
\showURL{%
\tempurl}


\bibitem[Li et~al\mbox{.}(2023)]%
        {alpa}
\bibfield{author}{\bibinfo{person}{Zhuohan Li}, \bibinfo{person}{Lianmin Zheng}, \bibinfo{person}{Yinmin Zhong}, \bibinfo{person}{Vincent Liu}, \bibinfo{person}{Ying Sheng}, \bibinfo{person}{Xin Jin}, \bibinfo{person}{Yanping Huang}, \bibinfo{person}{Zhifeng Chen}, \bibinfo{person}{Hao Zhang}, \bibinfo{person}{Joseph~E. Gonzalez}, {and} \bibinfo{person}{Ion Stoica}.} \bibinfo{year}{2023}\natexlab{}.
\newblock \showarticletitle{{AlpaServe}: Statistical Multiplexing with Model Parallelism for Deep Learning Serving}. In \bibinfo{booktitle}{\emph{17th USENIX Symposium on Operating Systems Design and Implementation (OSDI 23)}}. \bibinfo{publisher}{USENIX Association}, \bibinfo{address}{Boston, MA}, \bibinfo{pages}{663--679}.
\newblock
\showISBNx{978-1-939133-34-2}
\urldef\tempurl%
\url{https://www.usenix.org/conference/osdi23/presentation/li-zhouhan}
\showURL{%
\tempurl}


\bibitem[Lin et~al\mbox{.}(2024)]%
        {parrot}
\bibfield{author}{\bibinfo{person}{Chaofan Lin}, \bibinfo{person}{Zhenhua Han}, \bibinfo{person}{Chengruidong Zhang}, \bibinfo{person}{Yuqing Yang}, \bibinfo{person}{Fan Yang}, \bibinfo{person}{Chen Chen}, {and} \bibinfo{person}{Lili Qiu}.} \bibinfo{year}{2024}\natexlab{}.
\newblock \bibinfo{title}{Parrot: Efficient Serving of LLM-based Applications with Semantic Variable}.
\newblock
\showeprint[arxiv]{2405.19888}~[cs.LG]
\urldef\tempurl%
\url{https://arxiv.org/abs/2405.19888}
\showURL{%
\tempurl}


\bibitem[Liu et~al\mbox{.}(2023a)]%
        {llava}
\bibfield{author}{\bibinfo{person}{Haotian Liu}, \bibinfo{person}{Chunyuan Lin}, \bibinfo{person}{Qingyang Li}, \bibinfo{person}{Zhewei Hu}, \bibinfo{person}{Yong Wang}, \bibinfo{person}{Cho-Jui Hsieh}, {and} \bibinfo{person}{Jason Lee}.} \bibinfo{year}{2023}\natexlab{a}.
\newblock \showarticletitle{LLaVA: Large Language and Vision Assistant}.
\newblock \bibinfo{journal}{\emph{arXiv preprint arXiv:2304.08485}} (\bibinfo{year}{2023}).
\newblock


\bibitem[Liu et~al\mbox{.}(2023b)]%
        {netreduce}
\bibfield{author}{\bibinfo{person}{Shuo Liu}, \bibinfo{person}{Qiaoling Wang}, \bibinfo{person}{Junyi Zhang}, \bibinfo{person}{Wenfei Wu}, \bibinfo{person}{Qinliang Lin}, \bibinfo{person}{Yao Liu}, \bibinfo{person}{Meng Xu}, \bibinfo{person}{Marco Canini}, \bibinfo{person}{Ray C.~C. Cheung}, {and} \bibinfo{person}{Jianfei He}.} \bibinfo{year}{2023}\natexlab{b}.
\newblock \showarticletitle{In-Network Aggregation with Transport Transparency for Distributed Training}. In \bibinfo{booktitle}{\emph{Proceedings of the 28th ACM International Conference on Architectural Support for Programming Languages and Operating Systems, Volume 3}} (Vancouver, BC, Canada) \emph{(\bibinfo{series}{ASPLOS 2023})}. \bibinfo{publisher}{Association for Computing Machinery}, \bibinfo{address}{New York, NY, USA}, \bibinfo{pages}{376–391}.
\newblock
\showISBNx{9781450399180}
\urldef\tempurl%
\url{https://doi.org/10.1145/3582016.3582037}
\showDOI{\tempurl}


\bibitem[Mannam et~al\mbox{.}(2021)]%
        {campus-shuttle}
\bibfield{author}{\bibinfo{person}{Naga Praveen~Babu Mannam}, \bibinfo{person}{Basa Sidvik}, {and} \bibinfo{person}{P. Rajalakshmi}.} \bibinfo{year}{2021}\natexlab{}.
\newblock \showarticletitle{Powering Prediction of an Autonomous Campus Shuttle using {CFD}}. In \bibinfo{booktitle}{\emph{2021 International Conference on Smart Generation Computing, Communication and Networking (SMART GENCON)}}. \bibinfo{pages}{1--6}.
\newblock
\urldef\tempurl%
\url{https://doi.org/10.1109/SMARTGENCON51891.2021.9645802}
\showDOI{\tempurl}


\bibitem[Patel et~al\mbox{.}(2024)]%
        {splitwise}
\bibfield{author}{\bibinfo{person}{Pratyush Patel}, \bibinfo{person}{Esha Choukse}, \bibinfo{person}{Chaojie Zhang}, \bibinfo{person}{Aashaka Shah}, \bibinfo{person}{Íñigo Goiri}, \bibinfo{person}{Saeed Maleki}, {and} \bibinfo{person}{Ricardo Bianchini}.} \bibinfo{year}{2024}\natexlab{}.
\newblock \bibinfo{title}{Splitwise: Efficient generative LLM inference using phase splitting}.
\newblock
\showeprint[arxiv]{2311.18677}~[cs.AR]
\urldef\tempurl%
\url{https://arxiv.org/abs/2311.18677}
\showURL{%
\tempurl}


\bibitem[PyTorch(2023)]%
        {torch-serve}
\bibfield{author}{\bibinfo{person}{PyTorch}.} \bibinfo{year}{2023}\natexlab{}.
\newblock \bibinfo{title}{{TorchServe}}.
\newblock \bibinfo{howpublished}{\url{https://pytorch.org/serve/}}.
\newblock


\bibitem[Qin et~al\mbox{.}(2024)]%
        {mooncake}
\bibfield{author}{\bibinfo{person}{Ruoyu Qin}, \bibinfo{person}{Zheming Li}, \bibinfo{person}{Weiran He}, \bibinfo{person}{Mingxing Zhang}, \bibinfo{person}{Yongwei Wu}, \bibinfo{person}{Weimin Zheng}, {and} \bibinfo{person}{Xinran Xu}.} \bibinfo{year}{2024}\natexlab{}.
\newblock \bibinfo{title}{Mooncake: A KVCache-centric Disaggregated Architecture for LLM Serving}.
\newblock
\showeprint[arxiv]{2407.00079}~[cs.DC]
\urldef\tempurl%
\url{https://arxiv.org/abs/2407.00079}
\showURL{%
\tempurl}


\bibitem[Radford et~al\mbox{.}(2021)]%
        {clip}
\bibfield{author}{\bibinfo{person}{Alec Radford}, \bibinfo{person}{Jong~Wook Kim}, \bibinfo{person}{Chris Hallacy}, \bibinfo{person}{Aditya Ramesh}, \bibinfo{person}{Gabriel Goh}, \bibinfo{person}{Sandhini Agarwal}, \bibinfo{person}{Girish Sastry}, \bibinfo{person}{Amanda Askell}, \bibinfo{person}{Pamela Mishkin}, \bibinfo{person}{Jack Clark}, \bibinfo{person}{Gretchen Krueger}, {and} \bibinfo{person}{Ilya Sutskever}.} \bibinfo{year}{2021}\natexlab{}.
\newblock \showarticletitle{Learning Transferable Visual Models From Natural Language Supervision}.
\newblock \bibinfo{journal}{\emph{arXiv preprint arXiv:2103.00020}} (\bibinfo{year}{2021}).
\newblock


\bibitem[Ray et~al\mbox{.}(2024)]%
        {ragserve}
\bibfield{author}{\bibinfo{person}{Siddhant Ray}, \bibinfo{person}{Rui Pan}, \bibinfo{person}{Zhuohan Gu}, \bibinfo{person}{Kuntai Du}, \bibinfo{person}{Ganesh Ananthanarayanan}, \bibinfo{person}{Ravi Netravali}, {and} \bibinfo{person}{Junchen Jiang}.} \bibinfo{year}{2024}\natexlab{}.
\newblock \bibinfo{title}{RAGServe: Fast Quality-Aware RAG Systems with Configuration Adaptation}.
\newblock
\showeprint[arxiv]{2412.10543}~[cs.LG]
\urldef\tempurl%
\url{https://arxiv.org/abs/2412.10543}
\showURL{%
\tempurl}


\bibitem[Reusser(2024)]%
        {tabular}
\bibfield{author}{\bibinfo{person}{Fredy Reusser}.} \bibinfo{year}{2024}\natexlab{}.
\newblock \bibinfo{title}{Tabular Learning: Encoding for Entity and Context Embeddings}.
\newblock
\showeprint[arxiv]{2403.19405}~[cs.LG]
\urldef\tempurl%
\url{https://arxiv.org/abs/2403.19405}
\showURL{%
\tempurl}


\bibitem[Roberts(1978)]%
        {packetswitch}
\bibfield{author}{\bibinfo{person}{L.G. Roberts}.} \bibinfo{year}{1978}\natexlab{}.
\newblock \showarticletitle{The evolution of packet switching}.
\newblock \bibinfo{journal}{\emph{Proc. IEEE}} \bibinfo{volume}{66}, \bibinfo{number}{11} (\bibinfo{year}{1978}), \bibinfo{pages}{1307--1313}.
\newblock
\urldef\tempurl%
\url{https://doi.org/10.1109/PROC.1978.11141}
\showDOI{\tempurl}


\bibitem[{Romil Bhardwaj, Zhengxu Xia, Ganesh Ananthanarayanan, Junchen Jiang, Yuanchao Shu, Nikolaos Karianakis, Kevin Hsieh, Paramvir Bahl, Ion Stoica}(2022)]%
        {ekya}
\bibfield{author}{\bibinfo{person}{{Romil Bhardwaj, Zhengxu Xia, Ganesh Ananthanarayanan, Junchen Jiang, Yuanchao Shu, Nikolaos Karianakis, Kevin Hsieh, Paramvir Bahl, Ion Stoica}}.} \bibinfo{year}{2022}\natexlab{}.
\newblock \showarticletitle{Ekya: Continuous Learning of Video Analytics Models on Edge Compute Servers}. In \bibinfo{booktitle}{\emph{USENIX NSDI}}.
\newblock


\bibitem[Russakovsky et~al\mbox{.}(2015)]%
        {ImageNet}
\bibfield{author}{\bibinfo{person}{Olga Russakovsky}, \bibinfo{person}{Jia Deng}, \bibinfo{person}{Hao Su}, \bibinfo{person}{Jonathan Krause}, \bibinfo{person}{Sanjeev Satheesh}, \bibinfo{person}{Sean Ma}, \bibinfo{person}{Zhiheng Huang}, \bibinfo{person}{Andrej Karpathy}, \bibinfo{person}{Aditya Khosla}, \bibinfo{person}{Michael Bernstein}, \bibinfo{person}{Alexander~C. Berg}, {and} \bibinfo{person}{Li Fei-Fei}.} \bibinfo{year}{2015}\natexlab{}.
\newblock \showarticletitle{{ImageNet Large Scale Visual Recognition Challenge}}.
\newblock \bibinfo{journal}{\emph{International Journal of Computer Vision (IJCV)}} \bibinfo{volume}{115}, \bibinfo{number}{3} (\bibinfo{year}{2015}), \bibinfo{pages}{211--252}.
\newblock
\urldef\tempurl%
\url{https://doi.org/10.1007/s11263-015-0816-y}
\showDOI{\tempurl}


\bibitem[Sanvito et~al\mbox{.}(2018)]%
        {network-is-ai-accelerator}
\bibfield{author}{\bibinfo{person}{Davide Sanvito}, \bibinfo{person}{Giuseppe Siracusano}, {and} \bibinfo{person}{Roberto Bifulco}.} \bibinfo{year}{2018}\natexlab{}.
\newblock \showarticletitle{Can the Network be the AI Accelerator?}. In \bibinfo{booktitle}{\emph{Proceedings of the 2018 Morning Workshop on In-Network Computing}} (Budapest, Hungary) \emph{(\bibinfo{series}{NetCompute '18})}. \bibinfo{publisher}{Association for Computing Machinery}, \bibinfo{address}{New York, NY, USA}, \bibinfo{pages}{20–25}.
\newblock
\showISBNx{9781450359085}
\urldef\tempurl%
\url{https://doi.org/10.1145/3229591.3229594}
\showDOI{\tempurl}


\bibitem[Shen et~al\mbox{.}(2019)]%
        {nexus}
\bibfield{author}{\bibinfo{person}{Haichen Shen}, \bibinfo{person}{Lequn Chen}, \bibinfo{person}{Yuchen Jin}, \bibinfo{person}{Liangyu Zhao}, \bibinfo{person}{Bingyu Kong}, \bibinfo{person}{Matthai Philipose}, \bibinfo{person}{Arvind Krishnamurthy}, {and} \bibinfo{person}{Ravi Sundaram}.} \bibinfo{year}{2019}\natexlab{}.
\newblock \showarticletitle{Nexus: A GPU Cluster Engine for Accelerating DNN-Based Video Analysis}. In \bibinfo{booktitle}{\emph{Proceedings of the 27th ACM Symposium on Operating Systems Principles}} (Huntsville, Ontario, Canada) \emph{(\bibinfo{series}{SOSP '19})}. \bibinfo{publisher}{Association for Computing Machinery}, \bibinfo{address}{New York, NY, USA}, \bibinfo{pages}{322--337}.
\newblock
\showISBNx{9781450368735}
\urldef\tempurl%
\url{https://doi.org/10.1145/3341301.3359658}
\showDOI{\tempurl}


\bibitem[Shoeybi et~al\mbox{.}(2020)]%
        {megatron}
\bibfield{author}{\bibinfo{person}{Mohammad Shoeybi}, \bibinfo{person}{Mostofa Patwary}, \bibinfo{person}{Raul Puri}, \bibinfo{person}{Patrick LeGresley}, \bibinfo{person}{Jared Casper}, {and} \bibinfo{person}{Bryan Catanzaro}.} \bibinfo{year}{2020}\natexlab{}.
\newblock \bibinfo{title}{Megatron-LM: Training Multi-Billion Parameter Language Models Using Model Parallelism}.
\newblock
\showeprint[arxiv]{1909.08053}~[cs.CL]
\urldef\tempurl%
\url{https://arxiv.org/abs/1909.08053}
\showURL{%
\tempurl}


\bibitem[Simonyan and Zisserman(2014a)]%
        {twostream}
\bibfield{author}{\bibinfo{person}{Karen Simonyan} {and} \bibinfo{person}{Andrew Zisserman}.} \bibinfo{year}{2014}\natexlab{a}.
\newblock \bibinfo{title}{Two-Stream Convolutional Networks for Action Recognition in Videos}.
\newblock
\showeprint[arxiv]{1406.2199}~[cs.CV]
\urldef\tempurl%
\url{https://arxiv.org/abs/1406.2199}
\showURL{%
\tempurl}


\bibitem[Simonyan and Zisserman(2014b)]%
        {vgg}
\bibfield{author}{\bibinfo{person}{Karen Simonyan} {and} \bibinfo{person}{Andrew Zisserman}.} \bibinfo{year}{2014}\natexlab{b}.
\newblock \bibinfo{title}{Very Deep Convolutional Networks for Large-Scale Image Recognition}.
\newblock
\urldef\tempurl%
\url{https://doi.org/10.48550/ARXIV.1409.1556}
\showDOI{\tempurl}


\bibitem[Siracusano et~al\mbox{.}(2022)]%
        {n3ic}
\bibfield{author}{\bibinfo{person}{Giuseppe Siracusano}, \bibinfo{person}{Salvator Galea}, \bibinfo{person}{Davide Sanvito}, \bibinfo{person}{Mohammad Malekzadeh}, \bibinfo{person}{Gianni Antichi}, \bibinfo{person}{Paolo Costa}, \bibinfo{person}{Hamed Haddadi}, {and} \bibinfo{person}{Roberto Bifulco}.} \bibinfo{year}{2022}\natexlab{}.
\newblock \showarticletitle{Re-architecting Traffic Analysis with Neural Network Interface Cards}. In \bibinfo{booktitle}{\emph{19th USENIX Symposium on Networked Systems Design and Implementation (NSDI 22)}}. \bibinfo{publisher}{USENIX Association}, \bibinfo{address}{Renton, WA}, \bibinfo{pages}{513--533}.
\newblock
\showISBNx{978-1-939133-27-4}
\urldef\tempurl%
\url{https://www.usenix.org/conference/nsdi22/presentation/siracusano}
\showURL{%
\tempurl}


\bibitem[Swamy et~al\mbox{.}(2022)]%
        {taurus}
\bibfield{author}{\bibinfo{person}{Tushar Swamy}, \bibinfo{person}{Alexander Rucker}, \bibinfo{person}{Muhammad Shahbaz}, \bibinfo{person}{Ishan Gaur}, {and} \bibinfo{person}{Kunle Olukotun}.} \bibinfo{year}{2022}\natexlab{}.
\newblock \showarticletitle{Taurus: a data plane architecture for per-packet ML}. In \bibinfo{booktitle}{\emph{Proceedings of the 27th ACM International Conference on Architectural Support for Programming Languages and Operating Systems}} (Lausanne, Switzerland) \emph{(\bibinfo{series}{ASPLOS '22})}. \bibinfo{publisher}{Association for Computing Machinery}, \bibinfo{address}{New York, NY, USA}, \bibinfo{pages}{1099–1114}.
\newblock
\showISBNx{9781450392051}
\urldef\tempurl%
\url{https://doi.org/10.1145/3503222.3507726}
\showDOI{\tempurl}


\bibitem[Swamy et~al\mbox{.}(2023)]%
        {homunculus}
\bibfield{author}{\bibinfo{person}{Tushar Swamy}, \bibinfo{person}{Annus Zulfiqar}, \bibinfo{person}{Luigi Nardi}, \bibinfo{person}{Muhammad Shahbaz}, {and} \bibinfo{person}{Kunle Olukotun}.} \bibinfo{year}{2023}\natexlab{}.
\newblock \showarticletitle{Homunculus: Auto-Generating Efficient Data-Plane ML Pipelines for Datacenter Networks}. In \bibinfo{booktitle}{\emph{Proceedings of the 28th ACM International Conference on Architectural Support for Programming Languages and Operating Systems, Volume 3}} (Vancouver, BC, Canada) \emph{(\bibinfo{series}{ASPLOS 2023})}. \bibinfo{publisher}{Association for Computing Machinery}, \bibinfo{address}{New York, NY, USA}, \bibinfo{pages}{329–342}.
\newblock
\showISBNx{9781450399180}
\urldef\tempurl%
\url{https://doi.org/10.1145/3582016.3582022}
\showDOI{\tempurl}


\bibitem[{Vikranth Srivatsa, Zijian He, Reyna Abhyankar, Dongming Li, Yiying Zhang}(2024)]%
        {preeble}
\bibfield{author}{\bibinfo{person}{{Vikranth Srivatsa, Zijian He, Reyna Abhyankar, Dongming Li, Yiying Zhang}}.} \bibinfo{year}{2024}\natexlab{}.
\newblock \bibinfo{title}{{Preble: Efficient Distributed Prompt Scheduling for LLM Serving}}.
\newblock
\showeprint[arxiv]{2407.00023}~[cs.CV]
\urldef\tempurl%
\url{https://arxiv.org/abs/2407.00023}
\showURL{%
\tempurl}


\bibitem[{Viyom Mittal, Shixiong Qi, Ratnadeep Bhattacharya, Xiaosu Lyu, Junfeng Li, Sameer G Kulkarni, Dan Li, Jinho Hwang, KK Ramakrishnan, and Timothy Wood}(2021)]%
        {mu}
\bibfield{author}{\bibinfo{person}{{Viyom Mittal, Shixiong Qi, Ratnadeep Bhattacharya, Xiaosu Lyu, Junfeng Li, Sameer G Kulkarni, Dan Li, Jinho Hwang, KK Ramakrishnan, and Timothy Wood}}.} \bibinfo{year}{2021}\natexlab{}.
\newblock \showarticletitle{{Mu: an efficient, fair and responsive serverless framework for resource constrained edge clouds}}. In \bibinfo{booktitle}{\emph{SoCC}}.
\newblock


\bibitem[Wang et~al\mbox{.}(2024)]%
        {cachegen}
\bibfield{author}{\bibinfo{person}{Keith Wang}, \bibinfo{person}{Zhiyu Liu}, \bibinfo{person}{Xiang Zhao}, \bibinfo{person}{Haoyu Ma}, \bibinfo{person}{Kaixiang Chen}, {and} \bibinfo{person}{Zhiqiang Zhang}.} \bibinfo{year}{2024}\natexlab{}.
\newblock \showarticletitle{CacheGen: KV Cache Compression and Streaming for Fast Large Language Model Serving}. In \bibinfo{booktitle}{\emph{Proceedings of SIGCOMM 2024}}. \bibinfo{pages}{1571--1586}.
\newblock
\urldef\tempurl%
\url{https://cs.stanford.edu/~keithw/sigcomm2024/sigcomm24-final1571-acmpaginated.pdf}
\showURL{%
\tempurl}


\bibitem[Xiong and Zilberman(2019)]%
        {switches-dream-of-ml}
\bibfield{author}{\bibinfo{person}{Zhaoqi Xiong} {and} \bibinfo{person}{Noa Zilberman}.} \bibinfo{year}{2019}\natexlab{}.
\newblock \showarticletitle{Do Switches Dream of Machine Learning? Toward In-Network Classification}. In \bibinfo{booktitle}{\emph{Proceedings of the 18th ACM Workshop on Hot Topics in Networks}} (Princeton, NJ, USA) \emph{(\bibinfo{series}{HotNets '19})}. \bibinfo{publisher}{Association for Computing Machinery}, \bibinfo{address}{New York, NY, USA}, \bibinfo{pages}{25–33}.
\newblock
\showISBNx{9781450370202}
\urldef\tempurl%
\url{https://doi.org/10.1145/3365609.3365864}
\showDOI{\tempurl}


\bibitem[Yang et~al\mbox{.}(2023)]%
        {autogpt}
\bibfield{author}{\bibinfo{person}{Hui Yang}, \bibinfo{person}{Sifu Yue}, {and} \bibinfo{person}{Yunzhong He}.} \bibinfo{year}{2023}\natexlab{}.
\newblock \bibinfo{title}{Auto-GPT for Online Decision Making: Benchmarks and Additional Opinions}.
\newblock
\showeprint[arxiv]{2306.02224}~[cs.AI]
\urldef\tempurl%
\url{https://arxiv.org/abs/2306.02224}
\showURL{%
\tempurl}


\bibitem[Yao et~al\mbox{.}(2024)]%
        {cacheblend}
\bibfield{author}{\bibinfo{person}{Jiayi Yao}, \bibinfo{person}{Hanchen Li}, \bibinfo{person}{Yuhan Liu}, \bibinfo{person}{Siddhant Ray}, \bibinfo{person}{Yihua Cheng}, \bibinfo{person}{Qizheng Zhang}, \bibinfo{person}{Kuntai Du}, \bibinfo{person}{Shan Lu}, {and} \bibinfo{person}{Junchen Jiang}.} \bibinfo{year}{2024}\natexlab{}.
\newblock \bibinfo{title}{CacheBlend: Fast Large Language Model Serving for RAG with Cached Knowledge Fusion}.
\newblock
\showeprint[arxiv]{2405.16444}~[cs.LG]
\urldef\tempurl%
\url{https://arxiv.org/abs/2405.16444}
\showURL{%
\tempurl}


\bibitem[{Yinmin Zhong, Shengyu Liu, Junda Chen, Jianbo Hu, Yibo Zhu, Xuanzhe Liu, Xin Jin,Hao Zhang}(2024)]%
        {distserve}
\bibfield{author}{\bibinfo{person}{{Yinmin Zhong, Shengyu Liu, Junda Chen, Jianbo Hu, Yibo Zhu, Xuanzhe Liu, Xin Jin,Hao Zhang}}.} \bibinfo{year}{2024}\natexlab{}.
\newblock \bibinfo{title}{{DistServe: Disaggregating Prefill and Decoding for Goodput-optimized Large Language Model Serving}}.
\newblock


\bibitem[Zhang et~al\mbox{.}(2023)]%
        {shepherd}
\bibfield{author}{\bibinfo{person}{Hong Zhang}, \bibinfo{person}{Yupeng Tang}, \bibinfo{person}{Anurag Khandelwal}, {and} \bibinfo{person}{Ion Stoica}.} \bibinfo{year}{2023}\natexlab{}.
\newblock \showarticletitle{{SHEPHERD}: Serving {DNNs} in the Wild}. In \bibinfo{booktitle}{\emph{20th USENIX Symposium on Networked Systems Design and Implementation (NSDI 23)}}. \bibinfo{publisher}{USENIX Association}, \bibinfo{address}{Boston, MA}, \bibinfo{pages}{787--808}.
\newblock
\showISBNx{978-1-939133-33-5}
\urldef\tempurl%
\url{https://www.usenix.org/conference/nsdi23/presentation/zhang-hong}
\showURL{%
\tempurl}


\bibitem[Zhang et~al\mbox{.}(2024)]%
        {ralm}
\bibfield{author}{\bibinfo{person}{Zhihao Zhang}, \bibinfo{person}{Alan Zhu}, \bibinfo{person}{Lijie Yang}, \bibinfo{person}{Yihua Xu}, \bibinfo{person}{Lanting Li}, \bibinfo{person}{Phitchaya~Mangpo Phothilimthana}, {and} \bibinfo{person}{Zhihao Jia}.} \bibinfo{year}{2024}\natexlab{}.
\newblock \bibinfo{title}{Accelerating Retrieval-Augmented Language Model Serving with Speculation}.
\newblock
\showeprint[arxiv]{2401.14021}~[cs.LG]
\urldef\tempurl%
\url{https://arxiv.org/abs/2401.14021}
\showURL{%
\tempurl}


\bibitem[Zhou and Aryal(2024)]%
        {impute}
\bibfield{author}{\bibinfo{person}{Youran Zhou} {and} \bibinfo{person}{Sunil Aryal}.} \bibinfo{year}{2024}\natexlab{}.
\newblock \showarticletitle{A Comprehensive Review of Handling Missing Data: Exploring Special Missing Mechanisms}.
\newblock \bibinfo{journal}{\emph{arXiv preprint arXiv:2404.04905v1}} (\bibinfo{year}{2024}).
\newblock
\urldef\tempurl%
\url{https://arxiv.org/abs/2404.04905v1}
\showURL{%
\tempurl}
\newblock
\shownote{Contact: \texttt{echo.zhou@deakin.edu.au}, \texttt{sunil.aryal@deakin.edu.au}}.


\end{thebibliography}

\end{document}